\documentstyle[12pt,amsfonts]{article}
\newcommand{\tdot}[1]{ \stackrel{\cdot}{#1} }

\newcommand{\define}{ \stackrel{\triangle}{=} }

\def\be{\begin{equation}}
\def\ee{\end{equation}}
\def\ba{\begin{array}}
\def\ea{\end{array}}

\begin{document}
%---------------------------------------------------
\title{\bf Gravitational Shielding Effects in Gauge Theory of Gravity}
\author{{Ning Wu}
\thanks{email address: wuning@mail.ihep.ac.cn}
\\
\\
{\small Institute of High Energy Physics, P.O.Box 918-1,
Beijing 100039, P.R.China}}
\maketitle
\vskip 0.8in
%\noindent

~~

\vskip 0.4in
%\noindent

\begin{abstract}

In 1992, E.E.Podkletnov and R.Nieminen find that, 
under certain conditions, ceramic superconductor 
with composite structure has revealed weak shielding 
properties against gravitational force.  In classical 
Newton's theory of gravity and even in Einstein's 
general theory of gravity, there are no grounds of 
gravitational shielding effects. But in quantum
gauge theory of gravity, the gravitational shielding 
effects can be explained in a simple and natural
way. In quantum gauge 
theory of gravity, gravitational gauge interactions 
of complex scalar field can be formulated based on 
gauge principle. After spontaneous symmetry breaking,
if the vacuum of the complex scalar field is not 
stable and uniform, there will be a mass term of 
gravitational gauge field. When gravitational gauge 
field propagates in this unstable vacuum of the 
complex scalar field, it will decays exponentially, 
which is the nature of gravitational shielding 
effects. The mechanism of gravitational shielding 
effects is studied in this paper, and some main 
properties of gravitational shielding effects are 
discussed.

\end{abstract}
%-------------------------------------------------------

\newpage

\Roman{section}

\section{Introduction}
\setcounter{equation}{0}

In 1992, E.E.Podkletnov and R.Nieminen
find that, under certain conditions,
ceramic superconductor with composite structure has revealed
weak shielding properties against gravitational force\cite{01,02}.
The gravitational shielding effects are hard to be understood
in the traditional theory of gravity. The early attempt at
a theoretical explanations were proposed by G.Modanese.\cite{03}
In classical Newton's theory of gravity and even in Einstein's
general theory of gravity, there are no grounds of gravitational
shielding effects. So, what is the mechanism of gravitational
shielding effects? It is known that ordinary superconductor
reveals no shielding properties against gravitational force,
but in Podkletnov experiment, a specially prepared ceramic
superconductor can reveal a weak gravitational shielding
effect. Which properties of the specially  prepared
ceramic superconductor take a key role in making the
superconductor reveal weak gravitational shielding effects?  \\

It is well known that four kinds of fundamental interactions
in nature can be well described by gauge field theory, and gauge
principle is the common nature of all fundamental interactions.
Great achievements of gauge field theory in describing strong,
electromagnetic and weak interactions make us believe that
the prospective quantum theory for gravity should be
gauge field theory.
Gauge treatment of gravity was suggested immediately
after the gauge theory birth itself\cite{04,05,06,07,08,09}.
In 2001, a completely new quantum gauge theory of gravity
is proposed by N.Wu, which is the first perturbatively
renormalizable quantum gravity in the 4-dimensional
Minkowski space-time\cite{10,11}. In this approach,
gravity, like electromagnetic interactions and strong
interactions, is treated as a kind of physical interactions
in flat space-time, not space-time geometry, so our basic
physical space-time is always flat. In this new approach,
the unification of fundamental interactions can be
formulated in a simple and beautiful way\cite{12,13,14}.
If we use the mass generation mechanism which is proposed
in literature \cite{15}, we can propose a new
theory on gravity which contains massive graviton and
the introduction of massive graviton does not affect
the strict local gravitational gauge symmetry of the
Lagrangian and does not affect the traditional long-range
gravitational force\cite{16}. The existence of massive graviton
will help us to understand the possible origin of dark
energy and dark matter in the Universe. In a recent work,
it is found that the equation of motion of a classical
mass point in gravitational field given by gauge theory
of gravity is different from the geodesic equation of
general relativity. Combining this equation of motion of
a mass point with the perturbative solution of the
field equation of gravitational gauge field, a correct
theoretical expectation on classical tests of gravity
can be obtained\cite{19}. All these achievements make
us believe that gauge theory of gravity is a possible
prospective theory of quantum gravity. Based on this
belief, we try to use gauge theory of  gravity to
explain quantum effects of gravitational interactions,
such as phase effects in COW experiments\cite{18,1801,1802},
and we find that gauge theory of gravity
gives out correct prediction on it\cite{20}. In this paper,
the gravitational shielding effects are studied in the
framework of gauge theory of gravity. It is found that
the gravitational shielding effects in Podkletnov
experiments can be well explained by gauge theory
of gravity in a simple and natural way. The reason
why ordinary superconductor reveals no
gravitational shielding effects but the specially prepared
ceramic superconductor can reveal weak gravitational
shielding effects is also discussed. The nature of
gravitational shielding effects is studied. \\

\section{Gravitational Gauge Field}
\setcounter{equation}{0}

First, for the sake of integrity,
we give a simple introduction to gravitational
gauge theory and introduce some notations which is used
in this paper. Details on quantum gauge theory of gravity
can be found in literatures \cite{10,11}. According to discussions
in literature \cite{1101}, the symmetry for gravitational
interactions can not be local Lorentz symmetry, and
gauge principle tells us that the correct symmetry for
gravitational interactions should be local translation
symmetry, or gravitational gauge symmetry\cite{10,11}.
In gauge theory of gravity, the most fundamental
quantity is gravitational gauge field $C_{\mu}(x)$,
which is the gauge potential corresponding to gravitational
gauge symmetry. Gauge field $C_{\mu}(x)$ is a vector in
the corresponding Lie algebra, which, for the sake
of convenience, will be called gravitational Lie algebra
in this paper. So $C_{\mu}(x)$ can expanded as
\be \label{2.10}
C_{\mu}(x) = C_{\mu}^{\alpha} (x) \hat{P}_{\alpha},
~~~~~~(\mu, \alpha = 0,1,2,3)
\ee
where $C_{\mu}^{\alpha}(x)$ is the component field and
$\hat{P}_{\alpha} = -i \frac{\partial}{\partial x^{\alpha}}$
is the  generator of gravitational
gauge group, which satisfies
\be \label{2.10a}
[ \hat{P}_{\alpha} ~~,~~ \hat{P}_{\beta} ]=0.
\ee
Unlike the ordinary $SU(N)$ group, the commutability of the
generators of the gravitational gauge group does not mean that the
gravitational gauge group is an Abelian group. In fact,
the gravitational gauge group is a non-Ablelian group\cite{10,11}.
The gravitational gauge covariant derivative is given by
\be \label{2.9}
D_{\mu} = \partial_{\mu} - i g C_{\mu} (x)
= G_{\mu}^{\alpha} \partial_{\alpha},
\ee
where $g$ is the gravitational coupling constant and
$G$ is given by
\be \label{2.11}
G = (G_{\mu}^{\alpha}) = ( \delta_{\mu}^{\alpha} - g C_{\mu}^{\alpha} ).
\ee
Matrix $G$ is an important quantity in gauge theory
of gravity. Its inverse matrix is denoted as $G^{-1}$
\be \label{2.12}
G^{-1} = \frac{1}{I - gC} = (G^{-1 \mu}_{\alpha}).
\ee
Using matrix $G$ and $G^{-1}$, we can define two important
quantities
\be \label{2.13}
g^{\alpha \beta} = \eta^{\mu \nu}
G^{\alpha}_{\mu} G^{\beta}_{\nu},
\ee
\be \label{2.14}
g_{\alpha \beta} = \eta_{\mu \nu}
G_{\alpha}^{-1 \mu} G_{\beta}^{-1 \nu}.
\ee
\\

The  field strength of gravitational gauge field is defined by
\be \label{2.16}
F_{\mu\nu} \define \frac{1}{-ig} \lbrack D_{\mu}~~,~~D_{\nu} \rbrack.
\ee
Its explicit expression is
\be \label{2.17}
F_{\mu\nu}(x) = \partial_{\mu} C_{\nu} (x)
-\partial_{\nu} C_{\mu} (x)
- i g C_{\mu} (x) C_{\nu}(x)
+ i g C_{\nu} (x) C_{\mu}(x).
\ee
$F_{\mu\nu}$ is also a vector in gravitational Lie algebra,
\be \label{2.18}
F_{\mu\nu} (x) = F_{\mu\nu}^{\alpha}(x) \cdot \hat{P}_{\alpha},
\ee
where
\be \label{2.19}
F_{\mu\nu}^{\alpha} = \partial_{\mu} C_{\nu}^{\alpha}
-\partial_{\nu} C_{\mu}^{\alpha}
-  g C_{\mu}^{\beta} \partial_{\beta} C_{\nu}^{\alpha}
+  g C_{\nu}^{\beta} \partial_{\beta} C_{\mu}^{\alpha}.
\ee
\\

\section{Gravitational Interactions of Complex Scalar Field}
\setcounter{equation}{0}

In quantum theory, superconductive state, which is a macroscopic
quantum state formed by Bose condensation of Cooper pairs in
superconductivity, is described by a complex scalar field. So,
in order to understand gravitational shielding effects in Podkletnov
experiments, the gravitational interactions of complex scalar
field should be studied first\cite{1601}. In principle,
electromagnetic interactions are also involved in the Podkletnov
experiments, but they do not take an essential role in the
understanding of the gravitational shielding effects, so,
for the sake of simplicity, the electromagnetic interactions
are not considered in the present model.
\\

In gauge theory of gravity, the lagrangian that describes
the  gravitational gauge interactions of complex scalar
field is
\be
{\cal L}_0 =
- \eta^{\mu\nu} (D_{\mu} \phi) (D_{\nu} \phi)^* - V(\phi)
- \frac{1}{4} \eta^{\mu \rho} \eta^{\nu \sigma} g_{\alpha \beta }
F_{\mu \nu}^{\alpha} F_{\rho \sigma}^{\beta},
\label{3.1}
\ee
where $V(\phi)$ is the potential of scalar field
\be
V(\phi) = \frac{1}{2} \mu^2 | \phi |^2
+ \frac{\lambda}{4} |\phi|^4,
~~~~~~(\lambda > 0).
\label{3.2}
\ee
The full Lagrangian of the system is
\be
{\cal L} = J(C) \cdot {\cal L}_0 ,
\label{3.3}
\ee
where
\be
J(C) = \sqrt{- {\rm det} g_{\alpha \beta} },
\label{3.4}
\ee
and the corresponding action is
\be
S =  \int {\rm d}^4 x {\cal L}
=\int {\rm d}^4 x  ~~ \sqrt{- {\rm det} g_{\alpha \beta} }
\cdot  {\cal L}_0 .
\label{3.5}
\ee
It can be strict proved that this action has strict local
gravitational gauge symmetry, and therefore, this model
is also perturbatively renormalizable. \\

When $\mu^2>0$, the minimum of the potential $V(\phi)$ is
at origin $\phi=0$, so $\phi$ is the quantum field. From
the above action, we can deduce the equation of motion of
complex scalar field $\phi (x)$ and the field equations
of gravitational gauge field $C_{\mu}^{\alpha}$. we can
also calculate the inertial energy-momentum tensor of the
system and the gravitational energy-momentum tensor of the
system. Then, we can study gravitational interactions of
complex scalar field\cite{10,11,1601}. \\

\section{Spontaneously Symmetry Breaking}
\setcounter{equation}{0}

When $\mu^2 <0$, the origin $\phi = 0$ will become a local
maximum and the symmetry of the system will be spontaneously
broken. After symmetry breaking, the physical vacuum moves
to $\phi_0$
\be
| \phi_0 | =  \sqrt{-\frac{\mu^2}{\lambda}}.
\label{4.1}
\ee
$\phi_0$ is the vacuum of the complex scalar field, generally
speaking, it is required that $\phi_0$ is even distributed
in microscopic scale, so that there is no quantum excitation,
otherwise $\phi_0$ is not a classical state, but a quantum
state. But on the other hand, $|\phi_0|^2$ represents
Cooper pairs density in superconductor, and generally speaking,
Cooper pairs density is not a constant in macroscopic scale,
so $\phi_0$ can not be a constant in macroscopic scale. So,
in this paper, our basic hypothesis on the physical vacuum
$\phi_0$ is that it is a classical state, it is even
distributed or a constant in microscopic scale and it
is a function of space-time coordinates in macroscopic
scale. In a word,
\be
\phi_0 = \phi_0(x).
\label{4.2}
\ee
$\phi_0$ is a classical state, it has no quantum excitation,
though it is space-time dependent. After symmetry breaking,
the scalar field becomes
\be
\phi(x) = \varphi(x) + \phi_0.
\label{4.3}
\ee
$\varphi(x)$ is the quantum scalar field after symmetry
breaking, which represents the small perturbations around
physical vacuum $\phi_0$. \\

After symmetry breaking, lagrangian becomes
\be
\ba{rcl}
{\cal L}_0 & = &
- \eta^{\mu\nu} (D_{\mu} \varphi) (D_{\nu} \varphi)^*
- \eta^{\mu\nu} (D_{\mu} \varphi) (D_{\nu} \phi_0)^*
- \eta^{\mu\nu} (D_{\mu} \varphi)^*  (D_{\nu} \phi_0) \\
&&\\
&&
- \eta^{\mu\nu} (D_{\mu} \phi_0) (D_{\nu} \phi_0)^*
- V(\varphi)
- \frac{1}{4} \eta^{\mu \rho} \eta^{\nu \sigma} g_{\alpha \beta }
F_{\mu \nu}^{\alpha} F_{\rho \sigma}^{\beta},
\ea
\label{4.4}
\ee
where $V(\varphi)$ is the potential of scalar field
\be
V(\varphi)  =  \frac{\lambda}{4} |\varphi|^4
+ \frac{\lambda}{2} |\varphi|^2
(\varphi^* \phi_0 + \varphi \phi_0^*)
- \frac{\mu^2}{2} |\varphi|^2
 + \frac{\lambda}{4}
(\varphi^{* 2} \phi_0^2 + \varphi^2 \phi_0^{*2} )
- \frac{\mu^4}{4 \lambda}.
\label{4.5}
\ee
There are two important properties which could be seen from
the above lagrangian. Firstly, when the vacuum $\phi_0$
is space-time dependent, it can directly couple to
quantum field $\varphi$. In other words, in the place where
$\partial_{\mu} \phi_0$ does not vanish, vacuum state $\phi_0$
can directly excite quantum scalar state, so it can be
considered to be an external source of scalar field.
Secondly, when the vacuum $\phi_0$ is space-time dependent,
it will couple to gravitational gauge field. But this interaction
is completely in the classical level. In other words, it is also
an external source of gravitational gauge field.
\\

The field equation of gravitational gauge field in the
superconductor is
\be
\label{4.6}
\partial_{\mu} (\eta^{\mu\rho} \eta^{\nu\sigma} g_{\alpha\beta}
F_{\rho \sigma}^{\beta}  ) = - g T_{g \alpha}^{\nu},
\ee
where $T_{g \alpha}^{\nu}$ is the gravitational energy-momentum
tensor
\be
\ba{rcl}
T_{g \alpha}^{\nu} & = &
\eta^{\mu\nu} (D_{\mu} \varphi^*) \partial_{\alpha} \varphi
+ \eta^{\mu\nu} (D_{\mu} \phi^*_0) \partial_{\alpha} \varphi
+ \eta^{\mu\nu} (D_{\mu} \varphi) \partial_{\alpha} \varphi^*
+ \eta^{\mu\nu} (D_{\mu} \phi_0) \partial_{\alpha} \varphi^*  \\
&&\\
&& + \eta^{\mu\nu} (D_{\mu} \varphi^*) \partial_{\alpha} \phi_0
+ \eta^{\mu\nu} (D_{\mu} \phi_0^*) \partial_{\alpha} \phi_0
+ \eta^{\mu\nu} (D_{\mu} \varphi) \partial_{\alpha} \phi_0^*
+ \eta^{\mu\nu} (D_{\mu} \phi_0) \partial_{\alpha} \phi_0^*  \\
&&\\
&& - \eta^{\mu\rho} \eta^{\nu\sigma} g_{\beta\gamma}
F_{\rho\sigma}^{\gamma} \partial_{\alpha} C_{\mu}^{\beta}
+ G^{-1 \nu}_{\alpha} {\cal L}_0
+ \eta^{\mu \rho} \eta^{\nu \sigma} g_{\alpha \beta}
G^{-1 \lambda}_{\gamma} (D_{\mu} C_{\lambda}^{\gamma})
F_{\rho \sigma}^{\beta}  \\
&&\\
&& - \frac{1}{2} \eta^{\mu \rho} \eta^{\lambda \sigma}
g_{\alpha \beta} G^{-1 \nu}_{\gamma}
F_{\mu\lambda}^{\beta} F_{\rho \sigma}^{\gamma}
- \eta^{\lambda \rho} \eta^{\nu \sigma}
\partial_{\mu} (g_{\alpha \beta}
C_{\lambda}^{\mu} F_{\rho\sigma}^{\beta} ).
\ea
\label{4.7}
\ee
In fact, when we deduce the field equation of gravitational
gauge field from least action principle, the original field
equation that we obtained is
\be
J(C) \cdot \partial_{\mu} (\eta^{\mu\rho} \eta^{\nu\sigma}
g_{\alpha\beta} F_{\rho \sigma}^{\beta}  )
= - g J(C) T_{g \alpha}^{\nu}.
\label{4.8}
\ee
Because $J(C)$ does not vanish, we can be eliminate it from
the above equation to obtain the field equation eq.(\ref{4.6}).
Now, in order to obtain correct gravitational shielding
effects, we need to start our discussions directly from
eq.(\ref{4.8}), which can be changed into another form
\be
\partial_{\mu} (\eta^{\mu\rho} \eta^{\nu\sigma}
g_{\alpha\beta} F_{\rho \sigma}^{\beta}  )
= - g J(C) T_{g \alpha}^{\nu}
+(1-J(C)) \partial_{\mu} (\eta^{\mu\rho} \eta^{\nu\sigma}
g_{\alpha\beta} F_{\rho \sigma}^{\beta}  ).
\label{4.9}
\ee
The field equation of complex scalar field is
\be
\ba{rcl}
\eta^{\mu\nu} D_{\mu} D_{\nu} \varphi - \frac{\mu^2}{2} \varphi
&=& - \eta^{\lambda \nu} (\partial_{\mu} G^{\mu}_{\lambda})
D_{\nu} \varphi - \eta^{\mu\nu} D_{\mu}D_{\nu} \phi_0  \\
&&\\
&& - \eta^{\lambda \nu} (\partial_{\mu} G_{\lambda}^{\mu})
D_{\nu} \phi_0
- g \eta^{\mu\nu} G^{-1 \kappa}_{\alpha}
(D_{\mu} C_{\kappa}^{\alpha}) D_{\nu} \varphi  \\
&&\\
&& - g \eta^{\mu\nu} G^{-1 \kappa}_{\alpha}
(D_{\mu} C_{\kappa}^{\alpha}) D_{\nu} \phi_0
- \frac{\lambda}{2} \varphi^* (\varphi + \phi_0)^2
- \frac{\lambda}{2} \varphi^2 \phi^*_0.
\ea
\label{4.10}
\ee
This field equation can be changed into another form
\be
\left (\eta^{\mu\nu} \partial_{\mu} \partial_{\nu}
- \frac{\mu^2}{2} \right ) \varphi =
-\eta^{\mu\nu} D_{\mu}D_{\nu} \phi_0
- \eta^{\lambda\nu} (\partial_{\mu} G^{\mu}_{\lambda})
D_{\nu} \phi_0
- g \eta^{\mu\nu} G^{-1 \kappa}_{\alpha}
(D_{\mu} C_{\kappa}^{\alpha}) D_{\nu} \phi_0 + \cdots.
\label{4.11}
\ee
From the right hand side of the above equation, we can see that
when the vacuum of the complex scalar field is not a constant,
it will become source of quantum scalar field $\varphi$. In other
words, inhomogeneous vacuum will excite quantum states. \\

\section{Gravitational Shielding Effect}
\setcounter{equation}{0}

After some complicated calculations, the field equation (\ref{4.9})
can be changed into the following form
\be
\partial_{\mu} (\eta^{\mu\rho} \eta^{\nu\sigma}
g_{\alpha\beta} F_{\rho \sigma}^{\beta}  )
= - g N^{\nu}_{\alpha}
+ g^2 M^{\nu\mu}_{\alpha \beta} C_{\mu}^{\beta} + \cdots,
\label{5.1}
\ee
where
\be
\ba{rcl}
N^{\nu}_{\alpha} & = &
\delta^{\nu}_{\alpha} \lbrack
- \eta^{\mu\lambda} (\partial_{\mu} \phi_0)
(\partial_{\lambda} \phi_0^* )+ \frac{\mu^4}{4 \lambda}
\rbrack  \\
&&\\
&& + \eta^{\mu\nu} (\partial_{\mu} \phi_0)
(\partial_{\alpha} \phi_0^* )
+ \eta^{\mu\nu} (\partial_{\mu} \phi_0^*)
(\partial_{\alpha} \phi_0 ),
\ea
\label{5.2}
\ee

\be
\ba{rcl}
M^{\nu\mu}_{\alpha\beta} & = &
(\delta^{\mu}_{\alpha} \delta^{\nu}_{\beta}
+  \delta^{\mu}_{\beta} \delta^{\nu}_{\alpha})
\lbrack \eta^{\rho\lambda} (\partial_{\rho} \phi_0)
(\partial_{\lambda} \phi_0^*)
- \frac{\mu^4}{4 \lambda}  \rbrack \\
&&\\
&& + (\eta^{\mu\nu} \delta^{\lambda}_{\beta}
- \eta^{\lambda \nu} \delta^{\mu}_{\beta})
\lbrack (\partial_{\lambda} \phi_0) (\partial_{\alpha} \phi_0^*)
+ (\partial_{\lambda} \phi_0^*) (\partial_{\alpha} \phi_0) \rbrack \\
&&\\
&& - \delta^{\nu}_{\alpha} \eta^{\mu\lambda}
\lbrack (\partial_{\lambda} \phi_0) (\partial_{\beta} \phi_0^*)
+ (\partial_{\lambda} \phi_0^*) (\partial_{\beta} \phi_0) \rbrack.
\ea
\label{5.3}
\ee
From above expressions, all terms in $N^{\nu}_{\alpha}$
and $M^{\nu\mu}_{\alpha\beta}$ are classical quantities,
so $N^{\nu}_{\alpha}$ and $M^{\nu\mu}_{\alpha\beta}$
themselves are classical quantities. Therefore, the first two
terms of the right hand side of eq.(\ref{5.1}) do not
represent interaction terms. The first term is the source
of gravity in superconductor, and the second term is
mass  term of gravitational gauge field. In fact, from
eq.(\ref{5.2}), we can see that $N^{\nu}_{\alpha}$
is just the energy-momentum tensor of the vacuum of complex
scalar field, which is the source of gravitational gauge
field. But, because the coupling constant of gravitational
interactions is extremely small and the total energy in
a superconductor is finite, the magnitude of total gravitational
gauge field generated by energy-momentum of the superconductor
is many orders smaller than that of the gravitational gauge
field generated by the earth. In experiments, we can neglect
the gravity generated by the superconductor itself. So,
$N^{\nu}_{\alpha}$ has no contribution to gravitational
shielding effects. We neglect this term for the moment. \\

Let's turn to the second term of the right hand side of
eq.(\ref{5.1}). It is a mass term, but it is not a
constant mass term, so it is a variable mass term or
local mass term. At different positions,
$M^{\nu\mu}_{\alpha\beta}$ has different values. It is
known that, in vacuum space, graviton is massless and
gravitational force are long range force. But in
superconductor, gravitational gauge field obtain a small
mass term, so in superconductor, gravitational force
decreased exponentially, which is the nature of gravitational
shielding effects. So, when gravitational field goes out
the superconductor, it will becomes much weaker than it goes
into the superconductor.
 \\

For Podkletnov experiments, the dominant component of earth's
gravitational gauge field is $C^0_0$, which corresponds
to classical Newton's gravity. So, in order to explain
gravitational shielding effects quantitatively, we need
to study the propagation of gravitational gauge field
$C_0^0$ in superconductor. For earth's gravitational
field $C_{\mu}^{\alpha}$, it is static, so we can set
all time derivative of $C_{\mu}^{\alpha}$ to zero.
From field equation (\ref{5.1}), the following field
equation can be obtained
\be
\nabla^2 C_0^0
=  g^2 M^{00}_{00} C_{0}^{0} + \cdots,
\label{5.4}
\ee
where
\be
M^{00}_{00} = 2 | \nabla \phi_0 |^2 + 2 V(\phi_0),
\label{5.7}
\ee
and the contribution from $N^0_0$ is neglected.
Now, the mass term becomes very simple form.
In ordinary superconductor, the Cooper pair density is almost
a constant and space gradient of $\phi_0$ is almost zero.
In this case,
\be
M^{00}_{00} = 2 V(\phi_0) = - \frac{\mu^4}{2 \lambda}.
\label{5.8}
\ee
After omitting contribution from $N^0_0$, the
field equation (\ref{5.4}) becomes
\be
\nabla^2 C^0_0 = - \frac{g^2 \mu^4}{2 \lambda} C^0_0.
\label{5.9}
\ee
Selecting spherical coordinate system. For the earth's gravitational
field, $C^0_0$ is approximately a function of $r$ coordinate. In this
case, the general solution of eq. (\ref{5.9}) is
\be
C^0_0 (r)=
\frac{c_1}{r} {\rm cos}
\left (\sqrt{\frac{g^2 \mu^4}{2 \lambda}} (r-r_0) \right)
+ \frac{c_2}{r} {\rm sin}
\left (\sqrt{\frac{g^2 \mu^4}{2 \lambda}} (r-r_0) \right),
\label{5.10}
\ee
where $r_0$ is the position of the lower surface of superconductor.
Generally speaking, $\frac{g^2 \mu^4}{2 \lambda}$ is a extremely
small quantity(later, we will give a rough estimation on it),
in a small region of ordinary superconductor, we have
\be
C^0_0 (r) \approx \frac{c_1}{r}.
\label{5.11}
\ee
That is, ordinary superconductor almost have no effects on
earth's gravitational field, or it shows no gravitational
shielding effects. But for Podkletnov experiment, $\phi_0$
is not a constant, $|\nabla \phi_0|^2$
is many orders larger than $2 V(\phi_0)$, so the dominant
contribution of $M^{00}_{00}$ is $2 |\nabla \phi_0|^2$ and
it becomes a positive quantity. In this case,
$M^{00}_{00}$ is not a constant. Denote $g^2 M^{00}_{00}$
by $m_g^2$
\be
m_g^2 = g^2 M^{00}_{00}= 2 g^2|\nabla \phi_0|^2,
\label{5.12}
\ee
and omit the influence from $N_0^0$, then the above field
equation (\ref{5.4}) becomes
\be
\nabla^2 C^0_0 = m_g^2 C_0^0.
\label{5.13}
\ee
The general solution of the above equation is
\be
C_0^0 (r) =
c_1 \frac{e^{-m_g (r-r_0)}}{r}
+ c_2 \frac{e^{m_g (r-r_0)}}{r}.
\label{5.14}
\ee
In Podkletnov experiment, $c_2$ vanishes and earth's gravitational
field decrease exponentially
\be
C_0^0 (r) =
c_1 \frac{e^{-m_g (r-r_0)}}{r}.
\label{5.15}
\ee
That is, the earth's gravitational field decrease exponentially
in inhomogeneous superconductor. Suppose that the thickness
of the gravitational shielding region is $\Delta r_0$ and $m_g$
denotes the average value of the mass of graviton in this region,
then the relative gravity loss in the upper surface of
 the superconducting disk is
\be
\varepsilon \approx m_g  \Delta r_0.
\label{5.16}
\ee

\section{Podkletnov Experiments }
\setcounter{equation}{0}

Now, let's get a rough estimation on gravitational shielding
effects in Podkletnov experiment. Because Newton's gravitational
constant is
\be
G_N = \frac{g^2}{ 4 \pi},
\label{6.1}
\ee
so
\be
m_g^2 =  8 \pi G_N |\nabla \phi_0|^2.
\label{6.2}
\ee
Obviously, increase space gradient of $\phi_0$ will increase
$m_g$, and therefore increase gravitational shielding effects.
According to this spirit, the key structure of the ceramic
superconductor disk in Podkletnov experiment is that is has
three zones with different crystal structure and the gravitational
shielding effects mainly come from the transition part of the disk,
which is consists of randomly oriented grains with typical
sizes between 5 and 15 $\mu m$. In the experiment, the upper
layer of the disk is superconducting, while the lower layer
is not. There is a transition region between the tow layers.
The upper part of the transition region is partially in
superconducting,  some part of the  transition region is
critical, while the lower part of the transition region is not
superconducting. Because  of the granular structure of the lower
part, the Cooper pair density is strongly inhomogeneous and the
space gradient of $\phi_0$ will be large. Besides, the
supercurrent is disturbed by a high frequency magnetic field,
which will increase the space gradient of $\phi_0$. Because
the supercurrent is only in the surface of superconductor, so
the space gradient of $\phi_0$ is mainly determined by
penetration depth $l_0$. Denote  the maximum
of $\phi_0$ as $\phi_{0m}$, then we get the following estimation
\be
|\nabla \phi_0|^2 \propto
\frac{|\phi_{om}|^2}{l_0^2} .
\label{6.3}
\ee
For a complex scalar field, its probability is
\be
\rho(x) = -i (\tdot{\phi^*} \phi - \phi^* \tdot{\phi}).
\label{6.4}
\ee
So, in superconductor, the Cooper pair density $\phi_0$ is
\be
\rho_0 = -i (\tdot{\phi^*}_0 \phi_0 - \phi^*_0 \tdot{\phi}_0).
\label{6.5}
\ee
For an ordinary supercurrent, suppose that its lifetime is $T_0$.
Generally speaking, the lifetime of supercurrent is several
years long. Then we can get the following estimation
\be
\tdot{\phi}_0 \propto \frac{\phi_{0m}}{T_0}.
\label{6.6}
\ee
Therefore,
\be
\rho_0 \propto \frac{|\phi_{0m}|^2}{T_0},
\label{6.7}
\ee
\be
|\nabla \phi_0|^2 \propto
\frac{\rho_0 T_0}{l_0^2} .
\label{6.8}
\ee
The approximate magnitude of the mass $m_g$ of graviton in
inhomogeneous supercurrent is
\be
m_g^2 \approx 8 \pi G_N \frac{\rho_0 T_0}{l_0^2} .
\label{6.9}
\ee
For a rough estimation, taking that
\be
\rho_0 \sim 10^{32} m^{-3},
\label{6.10}
\ee
\be
T_0 \sim 10~ {\rm year} \sim 3 \times 10^8 s,
\label{6.11}
\ee
and for Podkletnov experiment,
\be
l_0 \sim 10^{-8} m.
\label{6.12}
\ee
Using these rough estimations, we get
\be
m_g \sim 0.03 ~m^{-1}.
\label{6.13}
\ee
If the thickness of the disk is about 0.01 $m$, according to
eq.(\ref{5.16}), the relative gravity loss is about
\be
\varepsilon \sim 0.03 \%,
\label{6.14}
\ee
which quite close to the experimental value of 0.3--0.5 $\%$.
So, present model can semi-quantitatively explain gravitational
shielding effects in Podkletnov experiment. \\

So, now, we can understand the roles of some experimental techniques
in Podkletnov experiment. The role of the upper layer of the disk
is to keep high supercurrent density in the experiment. Gravitational
shielding effects come from the transition region of the disk. The
high frequency magnetic field is used to increase disturbance from
outside of the disk and therefore to increase inhomogeneity of the
suppercurrent and final gravitational shielding effects. Fast rotation
of the disk is also favorable for gravitational shielding effects, for
the effect of this rotation is that the direction of the disturbance
from magnetic field changes with high speed, which will increase
disturbance. We know that gravitational shielding effects come from
inhomogeneity of supercurrents, so ordinary superconductor, which
carries homogeneous supercurrents, has no gravitational shielding
properties.  \\

\section{Discussions }
\setcounter{equation}{0}

In this paper, the gravitational shielding effects are discussed
in the framework of quantum gauge theory of gravity. When
gravitational field propagate in an inhomogeneous classical vacuum,
graviton will obtain a small mass term. The gravitational force
transmitted by massive graviton will decrease exponentially.
So, in gauge theory
of gravity, the nature of gravitational shielding effects is
exponentially decreasing of gravitational field in inhomogeneous
vacuum of scalar field after spontaneous symmetry breaking. \\

In a meaning, the present model only gives a qualitative
explanation on gravitational shielding effects. Without a doubt,
gravitational shielding effect found by E.Podkletnov and R.Nieminen
is an important effect, which may
cause a scientific evolution in the near future, or even change
our life style. We can imagine that, in the near future,
people can "walk" freely in the sky without a wing.
In order to thoroughly understand
and completely grasp the law of gravitational shielding effects,
more experimental and theoretical study on it is needed. The
present model can give some predictions on gravitational
shielding effects:

\begin{enumerate}

\item In phase transition, ordinary supercurrent may reveal
gravitational shielding effects, though it reveals no gravitational
shielding effects in a stable superconducting state. This is
because the supercurrent is strongly unstable and inhomogeneous
in phase transition.

\item Inhomogeneous  ordinary superfluid can also
reveal gravitational shielding effects. In a word, under suitable
conditions, any kind of matter which has  superfluidity can reveal
gravitational shielding effects.

\item Quark-Gluon Plasma(QGP), which is a special state of matter,
can also reveal gravitational shielding effects. In QGP, inhomogeneous
QCD vacuum can make graviton obtain a small mass term. It is known
that in the early stage, our universe is in QGP state. When our
universe is in QGP state, long range gravity can be shielded by
inhomogeneous vacuum. So, local fluctuation of matter can cause
instability of gravity, which will enlarge fluctuations of matter,
which is favorable for the formation of galaxy\cite{21}.

\end{enumerate}


\begin{thebibliography}{99}

\bibitem{01} E.Podkletnov and R.Nieminen, Physica {\bf C 203}
    (1992) 271.
\bibitem{02} E. Podkletnov, {\it Weak gravitational
    shielding properties of  composite bulk $YBa_2 Cu_3 O_{7-x}$
    superconductor below 70 K under electro-magnetic field},
    report MSU-chem 95, (cond-mat/9701074).
\bibitem{03} G.Modanese, Europhys. Lett. {\bf 35} (1996) 413;
    G.Modanese, Phys. Rev. {\bf D 54} (1996) 5002;
    G.Modanese and J.Schnurer, {\it Possible quantum gravity
    effects in a charged Bose condensate under variable e.m.
    field}, report UTH-391/96, (gr-qc/9612022).
\bibitem{04} R.Utiyama, Phys.Rev.{\bf 101} (1956) 1597.
\bibitem{05} A.Brodsky, D.Ivanenko and G. Sokolik, JETPH
        41 (1961)1307; Acta Phys.Hung. {\bf 14} (1962) 21.
\bibitem{06} T.W.Kibble, J.Math.Phys. {\bf 2} (1961) 212.
\bibitem{07} D.Ivanenko and G.Sardanashvily, Phys.Rep. {\bf 94}
        (1983) 1.
\bibitem{08} F.W.Hehl, J.D.McCrea, E.W.Mielke and Y.Ne'eman
        Phys.Rep. {\bf 258} (1995) 1-171
\bibitem{09} F.W.Hehl, P. Von Der Heyde, G.D.Kerlick, J.M.Nester
        Rev.Mod.Phys. {\bf 48} (1976) 393-416
\bibitem{10} Ning Wu, Commun. Theor. Phys. (Beijing, China)
        {\bf 38} (2002): 151-156.
\bibitem{11} Ning Wu, {\it Gauge Theory of Gravity}, talk given
        at Meeting of the Division of Particles and Fields
        of American Physical Society at the College of
        William \& Mary(DPF2002), May 24-28, 2002,
        Williamsburg, Virginia, USA; hep-th/0109145, hep-th/0207254.
\bibitem{12} Ning Wu, Commun. Theor. Phys. (Beijing, China)
        {\bf 38} (2002): 322-326.
\bibitem{13} Ning Wu, Commun. Theor. Phys. (Beijing, China)
        {\bf 38} (2002): 455-460.
\bibitem{14} Ning Wu, Commun. Theor. Phys. (Beijing, China)
        {\bf 39} (2003): 561-568.
\bibitem{15}  Ning Wu, Commun.Theor.Phys., (Beijing, China)
        {\bf 36}(2001) 169-172.
\bibitem{16} Ning Wu, Commun. Theor. Phys. (Beijing, China)
        {\bf 39} (2003): 671-674.
\bibitem{19} Ning Wu, {\it Classical Tests of Gauge Theory of Gravity}
    (in preparation)
\bibitem{18} R.Colella, A.W.Overhauser and S.A.Werner,
        Phys.Rev.Lett. (1975):1472-1474.
\bibitem{1801} S.A.Werner, R.Colella and  A.W.Overhauser,
        Phys.Rev.Lett. (1975): 1053-1055.
\bibitem{1802} A.W.Overhauser and R.Colella,
        Phys.Rev.Lett. (1974):1237-1239.
\bibitem{20}    Ning Wu, {\it Gravitational Phase Effects in Gauge
    Theory of Garvity} (in preparation)
\bibitem{1101} Ning Wu, Tunan Ruan, {\it Problems on Foundations
        of General Relativity}, hep-th/0303258.
\bibitem{1601} Ning Wu, {\it Gravitational Gauge Interactions
        of scalar field},
    (has been accepted by Commun. Theor. Phys.)
\bibitem{21}    Ning Wu, {\it Gravitational Shielding Effect of
    Quark Gluon Plasma in Gauge Theory of Garvity} (in preparation)




\end{thebibliography}
\end{document}